\documentclass[aps,prb,preprint,superscriptaddress]{revtex4-1}
\usepackage{graphicx}
\usepackage{epstopdf}
\usepackage{amsmath}
\usepackage{mathrsfs}
\usepackage{accents}
\usepackage{xcolor}
\begin{document}
\title{Transport signatures of relativistic quantum scars in a graphene cavity}
\author {G. Q. Zhang}
\altaffiliation{These authors contributed equally to this work.}
\affiliation{Beijing Key Laboratory of Quantum Devices, Key Laboratory for the Physics and Chemistry of Nanodevices and Department of Electronics, Peking University, Beijing 100871, P. R. China}
\author {Xianzhang Chen}
\altaffiliation{These authors contributed equally to this work.}
\affiliation{School of Physical Science and Technology and Key Laboratory for Magnetism and Magnetic Materials of MOE,
Lanzhou University, Lanzhou, Gansu 730000, P. R. China}
\author {Li Lin}
\affiliation{Center for Nanochemistry, Beijing Science and Engineering Center for Nanocarbons, Beijing National Laboratory for Molecular Sciences, College of Chemistry and Molecular Engineering, Peking University, Beijing 100871, P. R. China}
\author {Hailin Peng}
\affiliation{Center for Nanochemistry, Beijing Science and Engineering Center for Nanocarbons, Beijing National Laboratory for Molecular Sciences, College of Chemistry and Molecular Engineering, Peking University, Beijing 100871, P. R. China}
\author {Zhongfan Liu}
\affiliation{Center for Nanochemistry, Beijing Science and Engineering Center for Nanocarbons, Beijing National Laboratory for Molecular Sciences, College of Chemistry and Molecular Engineering, Peking University, Beijing 100871, P. R. China}
\author {Liang Huang}
\email[Corresponding author: ]{huangl@lzu.edu.cn}
\affiliation{School of Physical Science and Technology and Key Laboratory for Magnetism and Magnetic Materials of MOE, Lanzhou University, Lanzhou, Gansu 730000, P. R. China}
\author {N. Kang}
\email[Corresponding author: ]{nkang@pku.edu.cn}
\affiliation{Beijing Key Laboratory of Quantum Devices, Key Laboratory for the Physics and Chemistry of Nanodevices and Department of Electronics, Peking University, Beijing 100871, P. R. China}
\author {H. Q. Xu}
\email[Corresponding author: ]{hqxu@pku.edu.cn}
\affiliation{Beijing Key Laboratory of Quantum Devices, Key Laboratory for the Physics and Chemistry of Nanodevices and Department of Electronics, Peking University, Beijing 100871, P. R. China}
\affiliation{Division of Solid State Physics, Lund University, Box 118, S-221 00 Lund, Sweden}
\date{\today}

\begin{abstract}
Wave function scars refer to localized complex patterns of enhanced wave function probability distributions in a quantum system. Existing experimental studies of wave function scars concentrate nearly exclusively on nonrelativistic quantum systems. Here we study a relativistic quantum cavity system realized by etching out from a graphene sheet by quantum transport measurements and theoretical calculations. The conductance of the graphene cavity has been measured as a function of the back gate voltage (or the Fermi energy) and the magnetic field applied perpendicular to the graphene sheet, and characteristic conductance contour patterns are observed in the measurements. In particular, two types of high conductance contour lines, i.e., straight and parabolic-like high conductance contour lines, are found in the measurements. The theoretical calculations are performed within the framework of tight-binding approach and Green's function formalism. Similar characteristic high conductance contour features as in the experiments are found in the calculations. The wave functions calculated at points selected along a straight conductance contour line are found to be dominated by a chain of scars of high probability distributions arranged as a necklace following the shape of cavity and the current density distributions calculated at these point are dominated by an overall vortex in the cavity. These characteristics are found to be insensitive to increasing magnetic field. However, the wave function probability distributions and the current density distributions calculated at points selected along a parabolic-like contour line show a clear dependence on increasing magnetic field, and the current density distributions at these points are characterized by the complex formation of several localized vortices in the cavity. Our work brings a new insight into  quantum chaos in relativistic particle systems and would greatly stimulate experimental and theoretical efforts towards this still emerging field. 
\end{abstract}
\maketitle

In mesoscopic chaotic structures, the wave functions of eigenstates can coalesce in particular coordinate space to form scars with enhanced probability distributions\cite{BirdIshibashiAoyagiEtAl1997,BirdFerryAkisEtAl1996, AkisFerryBird1996,ZozoulenkoBerggren1997,BirdAkisFerryEtAl1997,FerryAkisPivinJrEtAl1998, FerryAkisBird2004,FerryAkisBird2005, AkisFerryBird1997,AkisBirdFerry2002,Heller1984,Bogomolny1988,Berry1989,Sridhar1991,SteinStoeckmann1992,MarcusRimbergWesterveltEtAl1992,FromholdWilkinsonSheardEtAl1995,XuHuangLaiGrebogi2013}. Scars could be related to classical trajectories in semiclassical limit\cite{BirdFerryAkisEtAl1996,AkisFerryBird1996,FerryAkisBird2005} or result from a superposition of several regular eigenstates in a closed structure\cite{ZozoulenkoBerggren1997}. It is feasible to measure such quantum scars in an open quantum system via transport measurements. In spite of that coupling to an environment will wash out the characteristics of quantum states, a few eigenstates of the corresponding closed structure, which can effectively mediate the transport, can be selected and visualized by transport measurements \cite{Zurek1982,Zurek2003,FerryAkisBird2005,ZozoulenkoBerggren1997}. Scars have been experimentally observed in different systems, including microwave billiard\cite{Sridhar1991,SteinStoeckmann1992} and mesoscopic cavities \cite{MarcusRimbergWesterveltEtAl1992,FromholdWilkinsonSheardEtAl1995,BirdOlatonaNewburyEtAl1995,AkisFerryBird1997,BirdIshibashiAoyagiEtAl1997,AkisFerryBirdEtAl1999}. For example, at low magnetic fields, certain characteristic patterns in the conductance of an open mesoscopic system are shown to be related to the underlying energy spectrum of the corresponding closed structure\cite{PerssonPetterssonVonSydowEtAl1995,ZozoulenkoSachrajdaGouldEtAl1999,AkisFerryBirdEtAl1999}. 
But, so far, experimental studies of mesoscopic cavities are nearly exclusively performed in nonrelativistic systems described by the Schr\"{o}dinger equation with a quadratic energy dispersion in a corresponding infinite system. An interesting question is whether such scars can generally appear with new characteristics in relativistic quantum systems described by the Dirac equation which, for an infinite system, gives a linear energy dispersion. Theoretical works have predicted\cite{HuangLaiFerryEtAl2009,HuangLaiFerryEtAl2009a,FerryHuangYangEtAl2010,HuangLaiGrebogi2011,YingHuangLaiEtAl2012} unequivocal evidence of quantum scars in relativistic systems. But, on the experimental side, observation of quantum scars has only been achieved very recently in a mesoscopic graphene ring device using scanning gate technique.\cite{Cabosart2017}
Graphene is a two-dimensional material consisting of carbon atoms in the honeycomb lattice \cite{NovoselovGeimMorozovEtAl2004}. Graphene exhibits rich interesting physics properties \cite{ZhangTanStormerEtAl2005a,YangCohenLouie2007, McCannKechedzhiFal'koEtAl2006,ChenRosenblattBolotinEtAl2009,NetoGuineaPeresEtAl2009,BolotinSikesJiangEtAl2008}, such as linear energy dispersion in the vicinity of the Dirac point, chiral carriers of zero mass, and extremely high mobility. Thus, a graphene cavity is an excellent candidate for the study of quantum scars in a finite relativistic system.

In this work, we study the transport properties of an open graphene cavity and demonstrate the observation of transport signatures of quantum scars in the relativistic particle system. The cavity is made of chemical vapor deposition (CVD) grown graphene on a substrate of n-doped Si covered by a thin layer of SiO$_2$. The conductance of the graphene cavity is measured as a function of the Fermi energy and magnetic field. Characteristic patterns are found in the measurements and are analyzed in terms of the underlying energy spectra. The system is also studied by theoretical calculations based on Green's function method. The calculated conductance map (i.e., a plot of the conductance as a function of the Fermi energy and the magnetic field) is found to be in good agreement with the experiment. In both the measured and the calculated conductance maps, two distinct types of high conductance contour lines, i.e., straight and parabolic-like lines, are found. From the calculated local density of states (LDOS) and the current density distributions, we find that along a straight high conductance contour line the scar pattern remains almost unchanged with increasing magnetic field. The current in the cavity is found to circulate clockwise at a finite magnetic field, when the contour line has a positive slope, and anti-clockwise, when the contour line has a negative slope. On contrast, along a parabolic-like high conductance contour line, the charge density distribution displays a much more complex scar pattern and the current density distribution exhibits the formation of several local vortices. At zero magnetic field, the total effective areas enclosed by the vortices circulating in opposite directions are the same. However, with increasing magnetic field, this balance is broken and the difference in the total effective area enclosed by the vortices of opposite directions increases, leading to the observation of the parabolic-like high conductance contour lines in the conductance maps.


Our graphene cavity device was fabricated on a Si/SiO$_2$ substrate from monolayer graphene grown via CVD. The fabrication was started by transferring CVD grown graphene on a substrate of n-doped Si covered by a 300-nm-thick layer of SiO$_2$ \cite{LinLiRenEtAl2016}. After transferring, a standard 16-$\mu$m-long and 3-$\mu$m-wide Hall-bar structure with a cavity inside was fabricated by electron beam lithography (EBL) and reactive ion etching with oxygen plasma. Contacts were subsequently fabricated by an additional step of EBL and deposition of a bilayer of Ti/Au (10 nm/90 nm) by electron beam evaporation. Figure 1(a) displays a false-color atomic force microscope (AFM) image of the fabricated device and a schematic for the measurement setup. The highlighted dark red region is the graphene current channel. The small green regions are graphene flakes which are isolated by narrow trenches from the cavity structure. A zoom-in look of the cavity structure is shown in Figure 1(b). The graphene cavity structure is of octagonal shape with $\sim$1 $\mu$m in size and is connected to bulk graphene via two 400-nm-wide constrictions. The device also consists of a region without a fine cavity structure. This arrangement enables us to directly compare the transport measurements of the cavity structure with bulk graphene on the same device. The measurements were carried out by applying a constant current $I$ through the two most distant contacts, that is, the source and the drain, and recording voltage drop $V_1$ over the cavity structure and voltage drop $V_2$ over the  graphene bulk at the same time. The conductance of the cavity region is obtained as $G_1=I /V_1$ and the conductance of the reference bulk region is obtained as $G_2=I/V_2$. The magnetotransport measurements were performed in a $^{3}$He/$^{4}$He dilution refrigerator with magnetic field $B$ applied perpendicular to the graphene plane, using a standard ac lock-in technique (with a current bias of 10-100 nA at a frequency of 13 Hz). Before comparative studies of the cavity with the graphene bulk, the graphene sheet was characterized by standard Hall measurements. Figure 1(c) shows the measured Hall resistance $R_{xy}$ and longitudinal resistance $R_{xx}$ in the bulk graphene region at magnetic field $B=5$ T at 60 mK. Here, well developed quantized Hall plateaus are observed, demonstrating the high quality of the graphene \cite{ZhangTanStormerEtAl2005a}. The mobility extracted from the measurements is around 17,000 cm$^2$V$^{-1}$s$^{-1}$ at carrier density $n\approx1.0\times10^{11}$ cm$^2$ and the mean free path $l_e$ derived from semi-classical relation\cite{BolotinSikesJiangEtAl2008} $ l_e=(\hbar/e)\cdot\mu\cdot(\pi n)^{1/2}$ is about 100 nm.


Figures 2(a) and 2(b) show the conductance $G_2$ of the bulk graphene region and the conductance $G_1$ of the cavity region measured as a function of the magnetic field at different back gate voltages.  Here and after, the longitudinal bulk graphene conductance is denoted by $G_2$. In Figure 2(a), the feature of universal conductance fluctuations (UCFs)\cite{Datta1997}, i.e., aperiodic fluctuations with fluctuation amplitude $\delta g_{2D}\approx0.2e^{2}/h$ are observable. Through the theoretical prediction\cite{Datta1997} of $\delta g_{2D}\approx L_\varphi(W^{1/2}/L^{3/2})\delta g_0$, where $W$ and $L$ are the width and the length of the bulk graphene Hall bar, $L_\varphi$ is the phase coherence length, and $\delta g_0\sim e^2/h$, the phase coherence length in our graphene sample is estimated to be on the order of 1 $\mu$m,  which is consistent with previous experiments \cite{ChenBaeChialvoEtAl2010}. In Figure 2(b), instead of showing UCFs, the conductance curves are smoother. In addition, We can observe that conductance peaks (indicated by black arrows) appear at some particular magnetic fields. These characteristics remind us about the conductance enhancements via transport through the eigenstates of the corresponding closed system\cite{ZozoulenkoBerggren1997}. To explore further about the evolution of these conductance peaks we performed the conductance map measurements, i.e., the measurements of the conductance as a function of the back gate voltage and the magnetic field, for both the bulk and cavity regions. Figure 2(c) and 2(d) show the measured conductance maps (in a color scale) for the bulk and cavity regions in a back gate voltage window of $V_{BG}=$8 V to 12 V and a magnetic field window of $-0.2$ T to 0.2 T. The red regions in Figure 2(c) and 2(d) represent the regions with high conductance. It is seen in Figure 2(d) that the measured conductance map exhibits a characteristic ``monkey face'' pattern of enhanced conductance. However, the measured conductance map shown in Figure 2(c) displays a nearly structure-less distribution of the conductance. In early experiments, similar characteristic features as seen in Figure 2(d) were found and were linked to the underlying energy spectra of closed quantum cavities\cite{AkisFerryBirdEtAl1999,PerssonPetterssonVonSydowEtAl1995,AkisFerry1998,ZozoulenkoSachrajdaGouldEtAl1999,ZozoulenkoBerggren1997}. Furthermore, because the carriers injected through a quantum point contact will be in a collimated form \cite{MolenkampStaringBeenakkerEtAl1990,BeenakkerVanHouten1989,BeenakkerHouten1991,AkisFerryBird1996,ZozoulenkoSchusterBerggrenEtAl1997,AkisFerryBirdEtAl1999,BirdAkisFerryEtAl1997}, carriers can enter the cavity with a sufficiently large probability only in certain angles and thus only some particular eigenstates can be preferentially excited and can contribute to the carrier transport.

Figure 3(a) is the conductance map measured in the same magnetic field window of $-0.2$ to 0.2 T but a back gate voltage window of $V_{BG}$=0 V to 5.8 V. Red arrows beside the graph indicate the direction towards the Dirac point. Here, as we expected, the overall conductance is found to decrease with increasing back gate voltage, i.e., when the Fermi level moves towards the Dirac point. But, more importantly, the measured conductance map is found to exhibit complex contour patterns. To better understand these complex features,
full quantum-mechanical transport calculations were
performed for the graphene cavity structure within the Landauer formalism\cite{Landauer1970}, which relates the zero-temperature two-terminal conductance $G$ of the device to the transmission coefficient $T$ in the form of $G=\frac{2e^2}{h}T$. The transmission coefficient $T$ were calculated in the Green's function scheme within the tight-binding framework\cite{LiLu2008,HuangLaiFerryEtAl2009,FerryHuangYangEtAl2010}, which we will only briefly describe here (for further details of the calculations, see Supporting Information). The cavity device can be split into three parts: left lead, cavity and right lead. The two leads are set to be semi-infinite to simulate the open boundaries \cite{Datta1997}. The Green's function of the device is given by $G_{D}(E)=(EI-H_{D}-\Sigma_{L}-\Sigma_{R})^{-1}$, where $\Sigma_{L}$ and $\Sigma_{R}$ are the self-energies caused by the left and right leads, and $H_D$ is the tight-binding Hamiltonian of the graphene cavity with hopping terms up to the third-nearest-neighbor atoms included. The hopping energies are 2.8 eV, 0.28 eV, and 0.07 eV for the nearest, the second-nearest, and the third-nearest neighbors, respectively \cite{Wallace1947,ReichMaultzschThomsenEtAl2002,KretininYuJalilEtAl2013}. The coupling matrices between the leads and the cavity, $\Gamma_{L}(E)$ and $\Gamma_{R}(E)$, are given in terms of self-energies $\Gamma_{L,R}=i(\Sigma_{L,R}-\Sigma_{L,R}^{\dagger})$. The transmission $T$ is given by $T(E)=\mathrm{Tr}(\Gamma_{L}G_{D}\Gamma_{R}G_{D}^{\dagger})$. The LDOS can be obtained by $\rho=-\frac{1}{\pi}\mathrm{Im}[\mathrm{diag}(G_D)]$.
The local current flow is given by
$J_{i\to j}=\frac{4e}{h}\mathrm{Im}[H_{D,ij}C_{ji}^{n}(E)]$ \cite{Datta1997},
where $C^{n}=G_D\Gamma^{L}G_D^{\dag}$ is the electron correlation function.

Figure 3(b) shows the results of the calculations, which clearly succeed in reproducing the main features of the experiment results shown in Figure 3(a). The satisfactory agreement between the experiment and the theory inspires us to get further understanding of the characteristic patterns of high conductance contour lines by taking a close comparison between Figure 3(a) and Figure 3(b). Let us focus on the regions marked by two dashed rectangles in Figure 3(a), which we label as regions I and II. We note that region I is closer to the Dirac point than region II.

Figure 4(a) is a close-up plot of the measurements in region I of Figure 3(a), while Figure 4(b) is a close-up plot of the calculations in the corresponding region shown in Figure 3(b).  Here, more featured high conductance contour lines are observable. Surprisingly, the patterns observed in the measurements and the calculations are still well matched. Both Figure 4(a) and Figure 4(b) show similar straight high conductance contour lines (see, e.g., the lines marked by yellow dashed lines A and B) and parabolic-like high contour lines (see, e.g., the lines marked by green dashed lines C). To get the physical insights into these characteristic high conductance contour lines, we have computed the LDOS and the current density distribution at a few selected points along the lines. In the tight-binding formulation (see Supporting Information for details), the LDOS provides the wave function probability distribution (or charge density distribution) contributed by all the states at energy $E$ and a given magnetic field $B$, while the current density distribution provides the information about current paths for carriers with energy $E$ to pass through the cavity at magnetic field $B$.

Figures 4(c) and 4(d) show the calculated LDOS and current density distribution at five selected points, denoted by $a_n$ with $n=$1, 2, 3, 4, and 5, along the straight high conductance contour line A in Figure 4(b). Red regions in Figure 4(c) correspond to the regions with high charge density probability distributions. Note that here the color scales in different panels are different. The patterns seen in Figure 4(c) are highly reminiscent of scars of enhanced wave function probabilities in coordinate space\cite{Heller1984,ZozoulenkoBerggren1997}. For example, at point $a_1$, the wave functions are highly localized to the regions close to the boundary of the cavity, looking as a chain of pearls (scars) arranged in a peanut shell like structure.  At point $a_2$ where a finite magnetic field is applied, the wave functions remain localized to the regions close to the boundary of the cavity. The same localization characteristics are also seen in the wave function probability distributions at points $a_3$ to $a_5$, although the scars become slightly smeared. In Figure 4(d), the corresponding current density distributions calculated at the same five selected points along the straight high conductance contour line A are plotted. Here it is seen that at zero magnetic field, i.e., at point $a_1$, the current density distribution is symmetric with respect to the horizontal axis (marked by a dot-dashed red line). On both the upper and the lower side of the axis, we see an overall current flow from the left to the right, although the several sharp current turns inside the cavity are observable. At finite magnetic fields, i.e., at points $a_2$ to $a_5$, the current density distributions are no longer symmetric with respect to the horizontal axis. Here on the upper side the current flows from the left to the right, while on the lower side the current flows from the right to the left, leading to the formation of an overall clockwise current vortex in the cavity. Note that here a net current passing through the cavity still remains to be from the left to the right. Note also that although we find the wave function probability distribution patterns at points $a_2$ to $a_5$ are very similar in Figure 4(c), their corresponding current density distributions shown in Figure 4(d) do exhibit small but noticeable differences. For example, although very similar current density distribution patterns are found at points $a_2$ and $a_3$ or at points $a_4$ and $a_5$, a small difference can be seen when the current density distribution patterns at points $a_3$ and $a_4$  are compared. This difference is most likely caused by the difference in mixing of the scarring states with other states, since the two points lie on the two sides of another high conductance contour line.

Similar localization characteristics have been found in the calculated wave function probability distributions and current density distributions at points $b_m$ with $m=$1, 2, 3, and 4 selected along high conductance contour line B. Here, the wave functions are again highly localized to form a chain of scars in the regions close to the boundary of the cavity and the current density distributions are seen to form an overall vortex in the cavity. However, it is interesting to note that the current vortex found at each of these points is to rotate anti-clockwise, in difference from the results obtained at points $a_2$ to $a_5$. This difference is consistent with the fact that high conductance contour line B has a negative slope, which is on contrast to line A (line A has a positive slope).

In a semiclassical description, along a high conductance contour lines, the energy of a state in the cavity at a magnetic field is approximately proportional to the magnetic flux penetrating through an effective area S enclosed in the current paths of an effective total length L, i.e., $E=E_0\pm (v_F e S/L)B$,\cite{YingHuangLaiEtAl2012} where the sign depends on the orientation of the local current circulating the magnetic flux (see Supporting Information for details). As we have shown above, with increasing magnetic field along a straight high conductance contour line, the wave function probability distributions and the current density distributions remain roughly the same. Thus, the effective area enclosed in the current vortex is approximately unchanged with increasing magnetic field. As a result, the energy of the states increases linearly with increasing magnetic field as seen in line A in Figures 4(a) and 4(b). The negative slope seen in line B is because here the current vortices are anti-clockwise and thus the magnetic flux penetrating through the effective area enclosed in each of these vortices carries an opposite sign.

Figures 5(a) and 5(b) show the calculated LDOS and current density distributions at selected points $c_k$ with $k=$1, 2, 3, and 4 along parabolic-like  high conductance contour line C shown in Figure 4(b). Again, the charge density distributions shown in Figures 5(a) are all symmetric with respect to the horizontal axis. But, such symmetry is found in the current density distribution only at zero magnetic field as seen in panel $c_1$ of Figure 5(b). However, when comparing to the results shown in Figure 4(c) to 4(f),  significant differences are found. First, high density spots localized in the middle of the cavity and  arranged as vertically elongated X patterns are found in the charge density distributions. Thus, no closed orbit-like structures are seen. Second, much more complex structures are seen in the current density distributions. In particular, several small current vortices are present in the current density distributions and are spread over the cavity. Third, at zero magnetic field, clockwise and anti-clockwise orientated current vortices are symmetrically localized in the cavity and the areas enclosed by all clockwise and all anti-clockwise oriented vortices are equal. But, with increasing magnetic field, the area enclosed by all vortices oriented in one direction (say clockwise) grows slowly and the area enclosed by all vortices oriented in the opposite direction (say anti-clockwise) shrinks. This difference in the area enclosed by the vortices of two different directions at a finite magnetic field is in contrast to the results shown in Figures 4(d) and 4(f), i.e., instead of being a constant, the effective circulating area $S$ in this case increases with increasing magnetic field, which could be the origin of the observed parabolic-like magnetic field dependence of the energy as revealed by the high conductance contour lines C shown in Figures 4(a) and 4(b).

Figure 6(a) shows a close-up plot of the measured conductance map in region II of Figure 3(a) and Figure 6(b) shows a plot of the calculated conductance map in the corresponding region. By comparison of the results shown in Figures 6(a)and 6(b), similar features can again be found in the measurements and calculations. Straight high conductance contour lines can be recognized and are marked by dashed yellow lines in the figures. Figures 6(c) and 6(d) show the calculated LDOS and the current density distributions at four selected points $d_l$ with $l=$1, 2, 3, and 4 along the yellow dashed line D shown in Figure 6(b). Here, as we have seen in Figures 4(c) to 4(f), the charge density distributions and the current density distributions shown in Figures 6(c) and 6(d) display similar ring-like orbit structures, and exhibit little changes with increasing magnetic field. This result is consistent with the linear dependence of the state energy on the magnetic field as we discussed above. However, comparing to the results calculated for the region I shown in Figure 4, here we can recognize clearly that additional current paths appear along the edges of the cavity. This might manifests the higher conductance observed in this far from Dirac point region.

In summary, we have studied the quantum transport properties of a relativistic quantum cavity. The cavity was made from a CVD grown graphene sheet on a Si/SiO$_2$ substrate. The low-temperature measurements of the conductance map, i.e., the conductance in the linear response regime as a function of the back gate voltage and the magnetic field applied perpendicular to the graphene plane, have been carried out for the cavity device. The complex characteristic features were found in the measured conductance map. To analyze the underlying physics revealed in these measurements, the graphene cavity device was modeled by a third-nearest-neighbor tight-binding Hamiltonian, and the conductance, charge density distribution and current density distribution were calculated based on Green's function formalism. The calculated conductance map exhibits similar complex characteristics as observed in the measurements. The calculated charge density distributions show the formation of scars and the current density distributions display the formation of complex vortices in the cavity. In particular, both straight and parabolic-like high conductance contour lines were found in the calculated and measured conductance maps. It has been found that along a straight high conductance contour line, the scar pattern remains almost unchanged with increasing magnetic field, while the circulating direction of the current in the cavity at a finite magnetic field is closely related to the slope of the contour line--it circulates clockwise when the contour line has a positive slope but anti-clockwise when the contour line has a negative slope. Here it should be emphasized that the straight high conductance contour line and the associated characteristics found in the scar pattern and the current density distribution are inherit to a relativistic quantum cavity. However, along a parabolic-like high conductance contour line, it has been found that the charge density distribution displays a complex scar pattern and the current density distribution exhibits the formation of several local vortices. Furthermore, although at zero magnetic field the total effective areas enclosed by the vortices circulating in opposite directions are the same, this balance is broken at a finite magnetic field and the difference in the total effective area enclosed by the vortices of opposite directions changes with increasing magnetic field. Such a parabolic-like high conductance contour line has been commonly observed for nonrelativistic quantum system at low Fermi energy. But here we show it can also been observed in a relativistic quantum cavity. We expect that our work would stimulate experimental and theoretical studies of quantum chaos in relativistic quantum systems.

\begin{center}
\noindent{\bf Acknowledgments}
\end{center}
We acknowledge financial supports by the Ministry of Science and Technology of China (MOST) through the National Key Research and Development Program of China (No. 2016YFA0300601 and 2017YFA0303304), the National Natural Science Foundation of China (Nos. 11874071, 11774005, 11775101, 91221202 and 91421303, 11374019), and the Swedish Research Council (VR).

\clearpage

\begin{figure}[t]
\begin{center}
\includegraphics[width=0.8\textwidth]{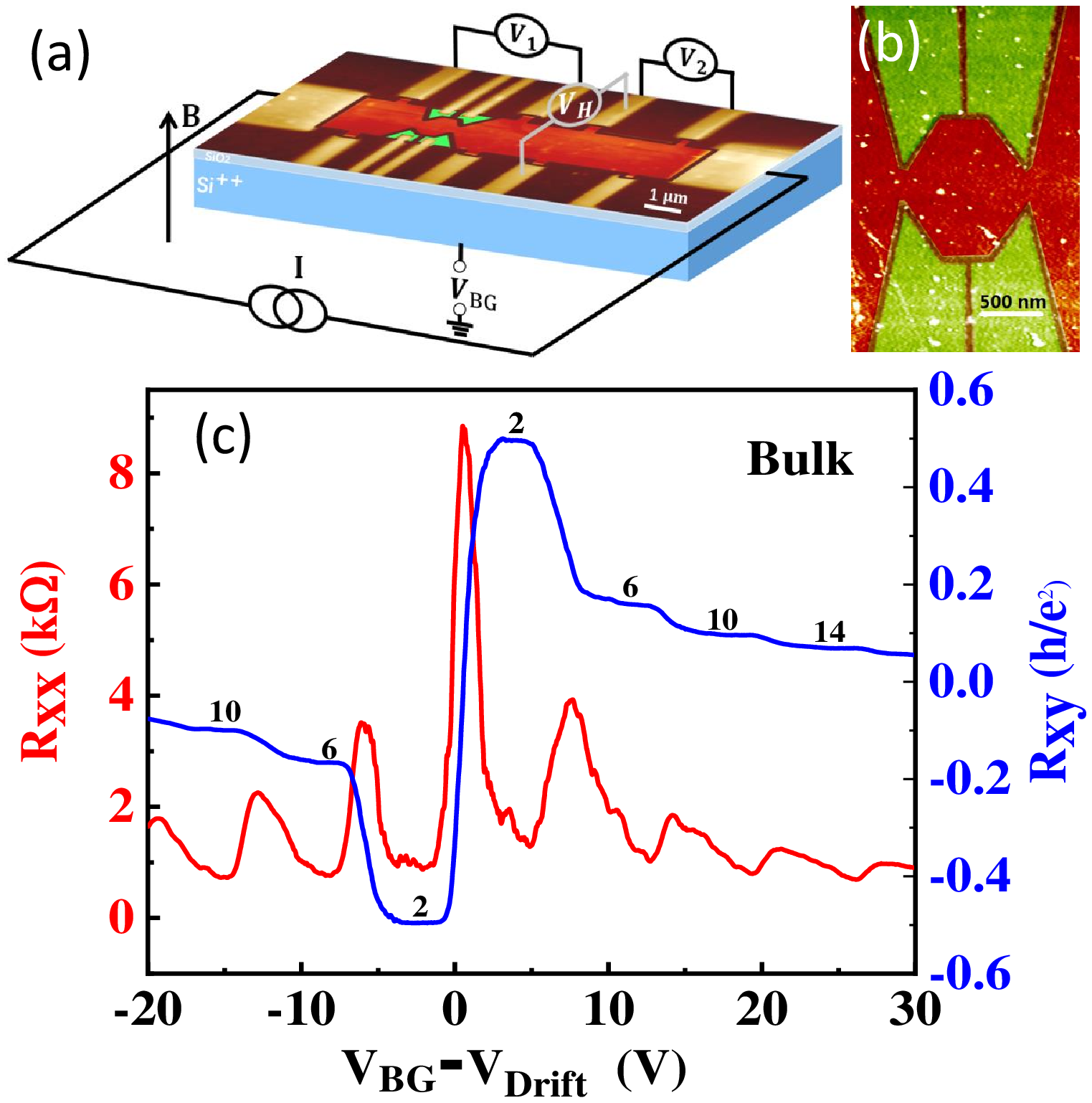}
\caption{(a) False-colored atomic force microscope (AFM) image  of the graphene device and schematics for the device structure and  measurement setup. Yellow parts are electrodes. Red region is graphene. Green highlighted regions are graphene pieces isolated from the cavity by etched trenches, which could be used as side gates but are not used in this work. The device is made on a Si/SiO$_2$ substrate which is used as a back gate.  $V_1$ and $V_2$ denote the voltage drops over the cavity and a graphene bulk region, respectively, and $V_H$ is the Hall voltage generated in the graphene bulk region. (b) Zoom-in AFM image of the cavity structure in (a). (c) Hall resistance $R_{xy}$ and longitudinal resistance $R_{xx}$ of the graphene bulk region measured at perpendicularly applied magnetic field $B=5$ T and temperature $T=60$ mK. $V_{BG}$ is the applied back gate voltage and $V_{\mbox{Drift}}$ is the drifting gate voltage of the Dirac point.}
\label{fig1}
\end{center}
\end{figure}
\clearpage

\begin{figure}[t]
\begin{center}
\includegraphics[width=\textwidth]{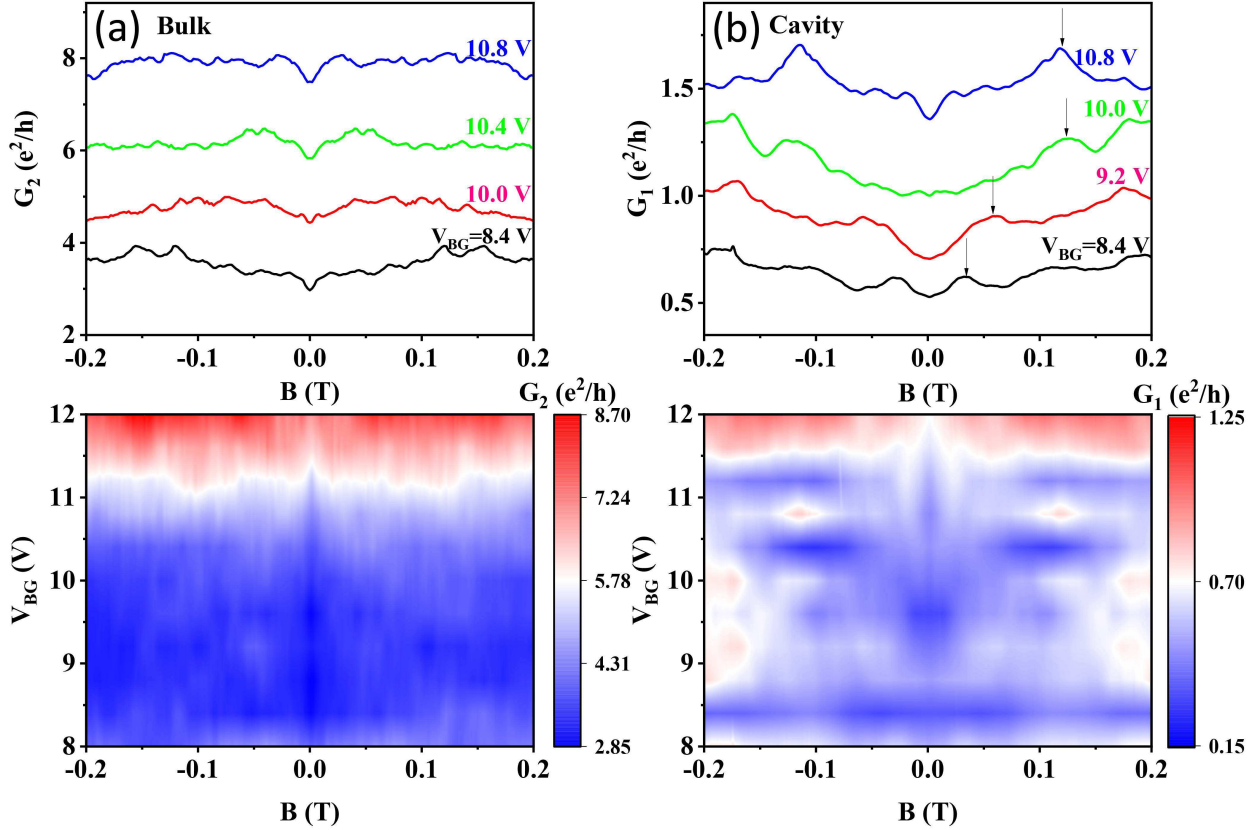}
\caption{(a) Longitudinal conductance of the graphene bulk region measured  as a function of the magnetic field $B$ at different back gate voltages $V_{BG}$ and temperature $T=60$ mK. Curves are successively vertically offset by
1${e^2}/h$ for clarity. (b) Conductance of the graphene cavity measured as a function of $B$ at different back gate voltages $V_{BG}$ and temperature $T=60$ mK.  Curves are successively vertically offset by 0.15${e^2}/h$. Black arrows in the figure denote the positions of conductance peaks which evolute with increasing back gate voltage. (c) Conductance map of the bulk graphene region, i.e., longitudinal conductance measured for the bulk graphene region as a function of $V_{BG}$ and $B$ at $T=60$ mK. (d) Conductance map of the graphene cavity at $T=60$ mK.}
\label{fig2}
\end{center}
\end{figure}
\clearpage

\begin{figure}[t]
\begin{center}
\includegraphics[width=13cm]{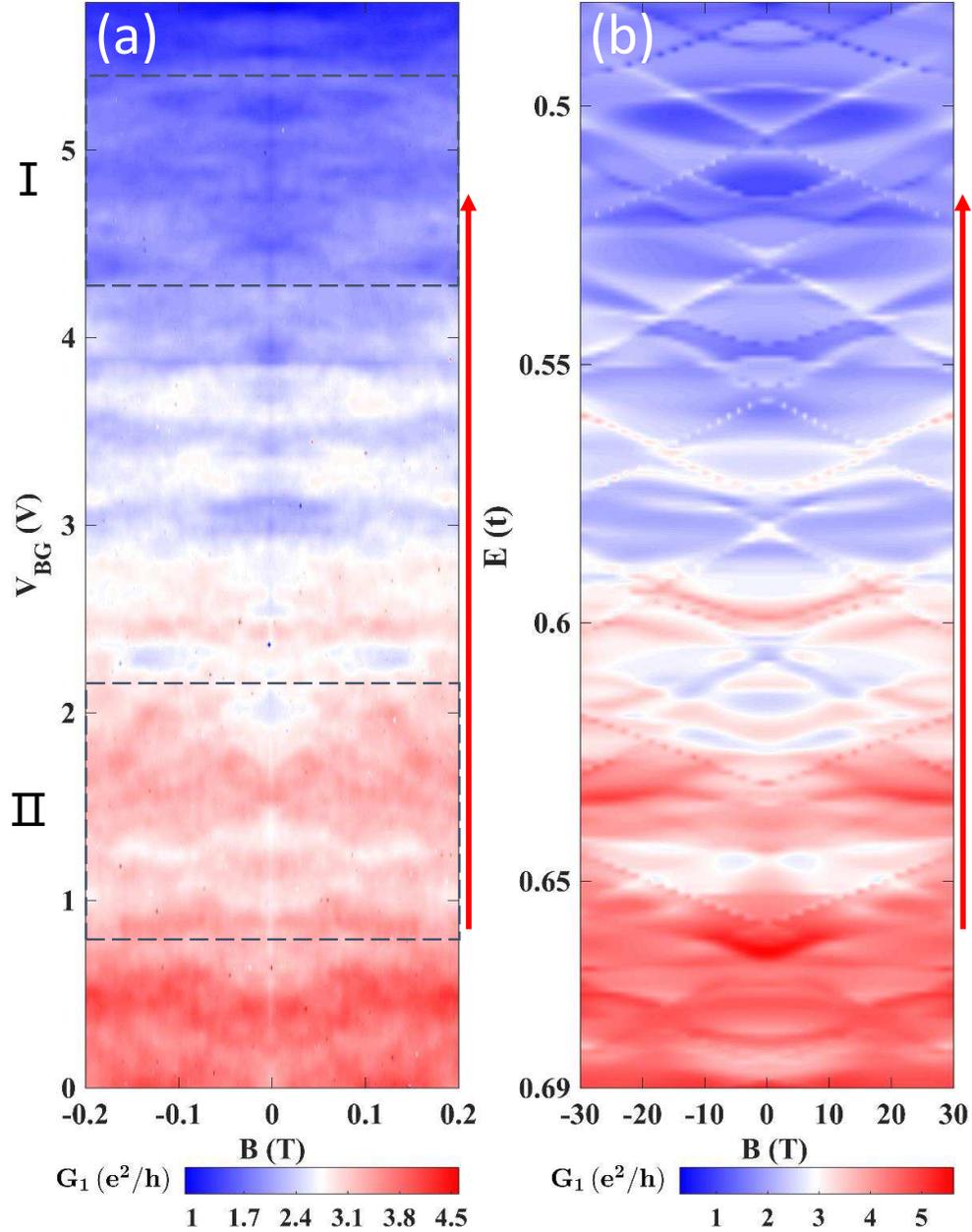}
\caption{(a) Conductance map of the graphene cavity measured at temperature $T=60$ mK over a large range of $V_{BG}$. (b) Simulated conductance map of the graphene cavity. Here, similar characteristic conductance patterns are found in the measurements and the calculations. The red arrows point to the direction towards the Dirac point.}
\label{fig3}
\end{center}
\end{figure}
\clearpage

\begin{figure}
\centering
\includegraphics[width=11cm]{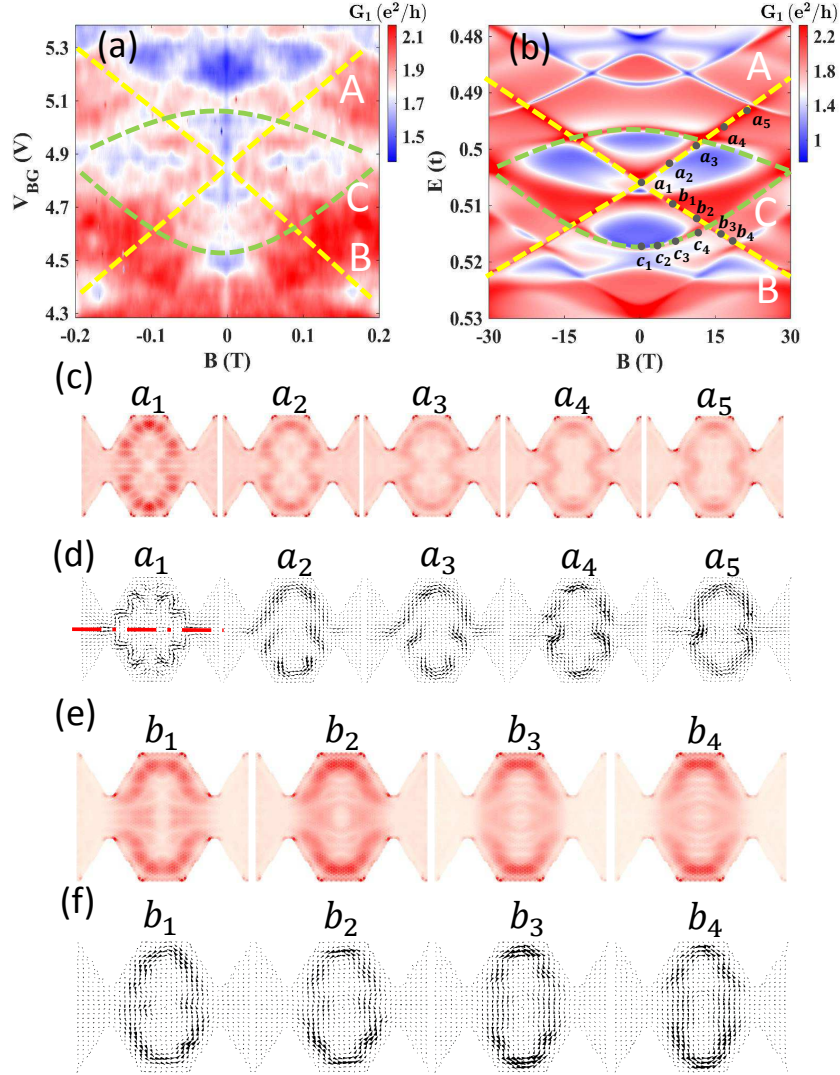}
\caption{(a) and (b) Zoom-in plot of the measurements shown in rectangular region I of Figure 3(a) and zoom-in plot of the calculations in the corresponding region shown in Figure 3(b). Straight (parabolic-like) high conductance contour lines are highlighted with yellow (green) dashed lines. (c) and (d) Calculated wave function probability distributions and current density distributions at points $a_n$, where $n=$1, 2, 3, 4, and 5, selected  along straight line A. Here, it is seen that the scar pattern does not show a significant change with increasing magnetic field and the current density distribution in each panel shows only one clockwise current vortex. (e) and (f) Calculated  wave function probability distributions and current density distributions at points $b_m$, where $m=$1, 2, 3, and 4, selected along straight contour line B. Similar characteristic features in the wave function probability distributions and the current density distributions as in (c) and (d) are observed, except that the current vortex seen in each current density distribution is anti-clockwise.}
\label{fig4}
\end{figure}

\begin{figure}[t]
\begin{center}
\includegraphics[width=\textwidth]{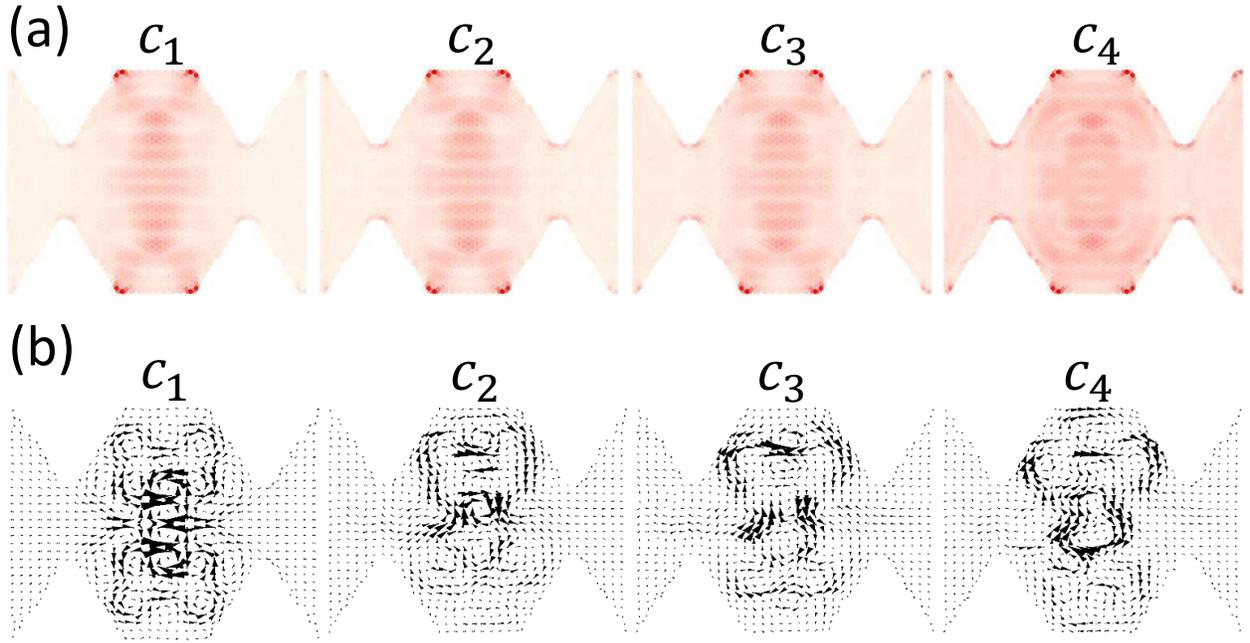}
\caption{(a) and (b) Calculated wave function probability distributions and current density distributions at points $c_k$, where $k=$1, 2, 3, and 4, selected along parabolic-like contour line C in figure 4(b). Here the scar distribution pattern shows sensitive change with change in magnetic field, and the current density distribution exhibits formation of a complex structure consisting of several localized clockwise and anti-clockwise current vortices.}
\label{fig5}
\end{center}
\end{figure}
\clearpage

\begin{figure}[t]
\begin{center}
\includegraphics[width=\textwidth]{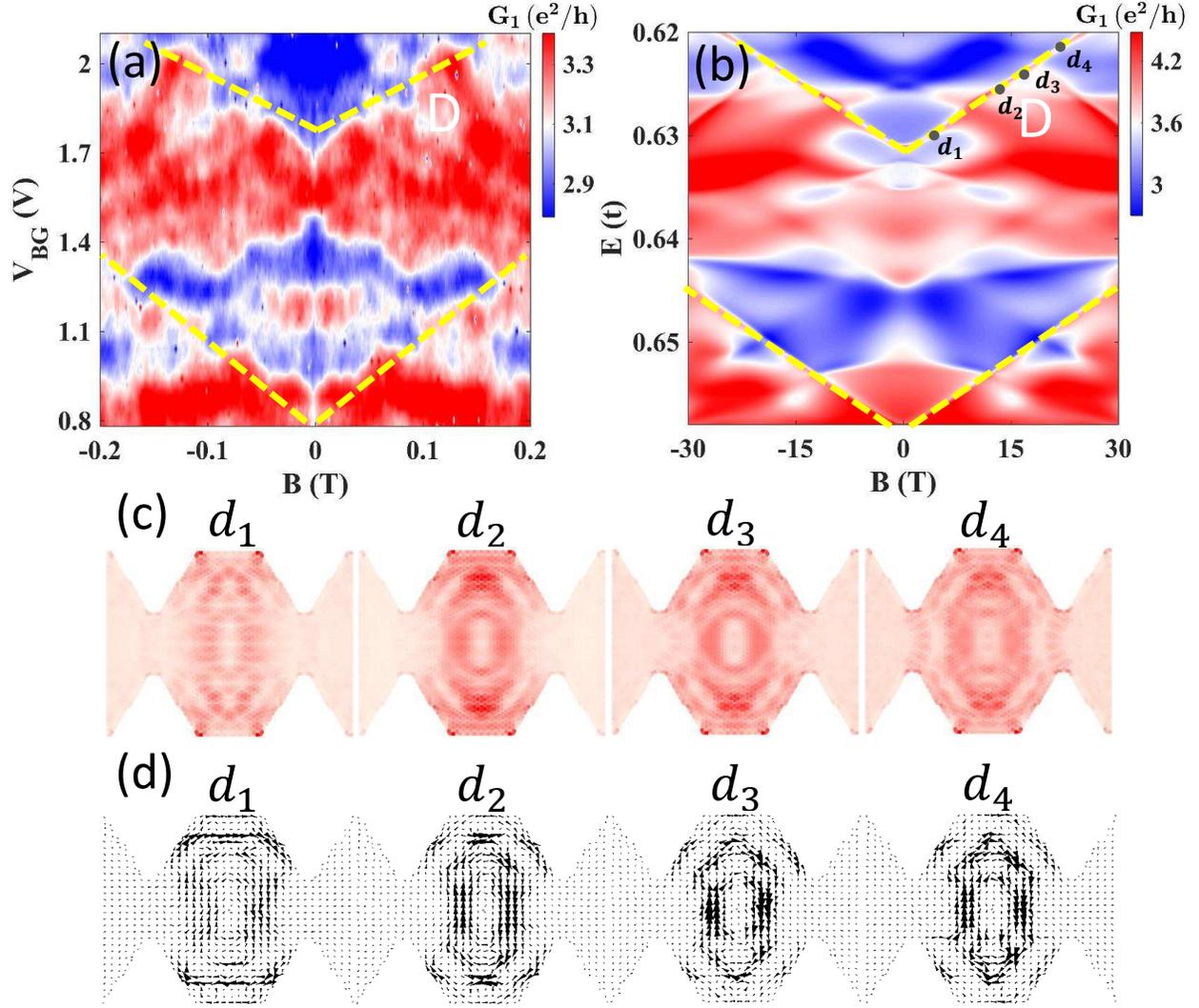}
\caption{(a) and (b) Zoom-in plot of the measurements shown in rectangular region II of Figure 3(a) and zoom-in plot of the calculations in the corresponding region shown in Figure 3(b). Note that region II is far from the Dirac point compared with region I. Yellow dashed lines highlights straight high  conductance contour lines observable in the measurements and the calculations. (c) and (d) Calculated wave function probability distributions and current density distributions at points $d_l$, where $l=$1, 2, 3, and 4, selected  along straight line D in (b). Similar characteristic features in the charge density distributions and the current density distributions as in Figure 4 are observed, except that a well-defined additional current path along the edge of the cavity is observable in each current density distribution panel.}
\label{fig6}
\end{center}
\end{figure}
\clearpage

\end{document}